\documentclass{sig-alternate}

\usepackage{enumitem}
\usepackage[hidelinks]{hyperref}
\usepackage{color}
\hypersetup{colorlinks=false}

\begin{document}


\makeatletter
\def\@copyrightspace{\relax}
\makeatother

\title{SCALL: Software Component Allocator for Heterogeneous Embedded Systems}
%
%
%
%
%

\numberofauthors{2} 
%
\author{
%
%
\alignauthor
Ivan \v{S}vogor\\
       \affaddr{University of Zagreb}\\
       \affaddr{Faculty of Organization and Informatics}\\
       \affaddr{Pavlinska 2, 42000 Vara\v{z}din, Croatia}\\
       \email{isvogor@foi.hr}
\alignauthor
Jan Carlson\\
       \affaddr{M\"{a}lardalen University}\\
       \affaddr{M\"{a}lardalen Real-Time Research Center}\\
       \affaddr{Box 883, 70123 V\"{a}ster\aa s , Sweden}\\
       \email{jan.carlson@mdh.se}
}


\maketitle

\begin{abstract}
Allocation of software components on a heterogeneous computing platform involves hard decisions; handling different types of computing units with specific processing paradigms and a number of software components which require specific resources. An allocation method which explores the design space to provide a system architect with deployment alternatives in an early design phase can have a significant impact on the utilization of underlying platform. In this paper we present SCALL, an early prototype tool which uses heuristics and AHP for weighted multi-objective design space exploration to support systems architects in complex allocation decisions in early design phases. 
\end{abstract}





\section{Introduction}


Heterogeneous computing has great potential in high\--per\-formance data processing, however with the benefits, it also brings new challenges for both hardware and software designers. These challenges are the subject of hardware--software codesign (HSCD), a field which investigates the concurrent design of hardware and software components in complex electronic systems. HSCD today seems to be skewed more towards hardware then software, however, with prevalence of off--the--shelf computing units this will change in future.
A literature review~\cite{teich2012hardware} by Teich has shown that software engineering is overlapping with hardware design and that HCSD techniques should be known to both hardware and software engineers involved in complex electronic systems, e.g. networked systems, automotive, avionics, etc. Tool supported specification, modeling, partitioning and synthesis of subsystems is of utmost importance for this field in order to enable efficient design of complex systems with tight non--functional constraints such as cost, performance, power, reliability, timing, etc.~\cite{6502240}.

The task of allocation (i.e. placing, mapping) is complex since it involves several research fields besides HSCD, e.g. adaptive computing systems, performance optimization, component based software engineering, etc. and by no means is it straightforward. For example, a system with $n$ components and $m$ computing units leaves software architects with $m^n$ possible allocations to choose from. This makes finding the best allocation laborious and even infeasible for large $n$ and $m$. Therefore, a tool which supports and automates this process is essential for aiding in architectural decision making. In this paper we present SCALL --- Software Component Allocator. It is an early prototype Eclipse plugin developed as part of our research on optimizing usage of heterogeneous computing platform by allocating software components.

\section{Objective}

In order to address issues described in the introduction we are using a component based approach which efficiently deals with static and dynamic configurations, variant explosions, scalability, reusability, etc. This results in a system composed of independent, more manageable, reusable and replaceable building blocks which enable various allocations based on the specification of extra--functional properties (energy consumption, processing time, memory requirements, etc.). Since different allocations of software components result in different system performance \cite{santinelli2010component}, our research goal is to construct a framework which optimizes the allocation of software components on a heterogeneous computing platform with respect to specified extra--functional requirements. Although similar tools exist, they largely fail to address heterogeneous platforms~\cite{Aleti2009, Malek2012, ccelik2013s}. 

\section{Fundamentals: the \\*mathematical model}
\label{sec:mmodel}

The basic elements of the (mathematical) model used by SCALL, are computing units, software components and the communication channels. The input for this model are matrices which quantify all the parameters necessary for making the allocation decision~\cite{Svogor2013}. The assumptions for the model are the following: a) the system is built out of atomic components, i.e. a component cannot be distributed over several computing units, b) each component can be deployed to any computing unit or a platform specific equivalent exist.

The system is a subject to a set of constraints which specify whether a particular component is supposed to be (or not be) allocated to a particular computing unit. To represent that kind of system the mathematical model uses five matrices: $\mathcal{T}$, $\mathcal{R}$, $\mathcal{C}$, $\mathcal{K}$ and $\mathcal{B}$ \footnote{Complete and precise mathematical definitions for all the matrices can be found in~\cite{Svogor2013}.}. 

\noindent $\mathcal{T}$ is a \textsl{resource consumption matrix} which specifies the resources required (i.e. consumed) by a component, since each component requires a certain amount of resources, which should be provided by the computing unit. This amount depends both on the kind of computing unit and the component. On the other hand, to specify resources which each computing unit can provide (e.g. available processing time, static memory, dynamic memory, energy etc.) we use the \textsl{computing unit resource matrix} $\mathcal{R}$.

\noindent The model considers communication in two aspects; a) hardware and b) software communication (as suggested in~\cite{Ristau2008}).  In one hand, hardware communication is limited by a physical constraint; bandwidth, represented by the \textsl{bandwidth matrix} $\mathcal{B}$. Additionally, since different computing units can use different communication channels and data types our model provides a \textsl{platform communication cost matrix} $\mathcal{C}$ which is a placeholder for expressing additional processing due to heterogeneous nature of such systems. 
\\*
\noindent On the other hand software communication can also cause overall system performance. Components which communicate intensively across computing units with high communication cost will have a larger impact on overall performance than ones communicating sporadically with lower data exchange. To express this constraint we define the \textsl{communication intensity matrix} $\mathcal{K}$. 

Finally, function  $p_i:S \to H$ is a \textsl{component mapping function} where $p_i=(p_1,...,p_n) \in  \mathcal{P}$ defines a particular allocation (from the set of possible allocations $\mathcal{P}$)  of components ($S$) to computing units($H$), with $i=1,...,m^n$. Elements of vector $p_i$ represent one mapping, i.e. allocation of components to the computing units. To find the \textsl{best} one, allocations are evaluated and compared using the cost function (1):

\vspace{-9mm}
\begin{center}
	\begin{large}
		\begin{equation}
		\label{eq:costfunction}
		w = \left(\sum\limits_{k=1}^l f_k  \sum\limits_{i=1}^n t_{i p_i k} + f_c \sum\limits_{i\leq j}k_{ij} \cdot c_{p_i p_j} \right) \cdot \rho \cdot \kappa ,
		\end{equation}  
	\end{large}
\end{center}

\vspace{-2mm}

\noindent where: 
$l$ is the number of resources used in matrix $\mathcal{T}$, 
$n$ is the number of components, 
$t$ is an element of a resource consumption matrix $\mathcal{T}$, 
$k$ is an element of a communication intensity matrix $\mathcal{K}$,
$b$ is an element of the bandwidth matrix $\mathcal{B}$,
$c$ is an element of the communication cost matrix $\mathcal{C}$,
$f$ is an element of a trade--off vector $F$.  
Given an architect's preference about importance of different resources from matrix $\mathcal{R}$, a trade--off vector $\mathcal{F}$ is calculated and it provides weights for different resources. To calculate $\mathcal{F}$ we are using Analytic Hierarchy Process (AHP~\cite{saaty1994fundamentals}).The multiplier $\rho$ and $\kappa$ are used to remove infeasible allocations. $\rho=0$ when the allocation uses more resources than it is available, effectively disregarding it. Similarly, multiplier $\kappa$ prevents an allocation in which the communication exceeds the available bandwidth. In all other cases both parameters are set to $1$. 

\noindent To find the optimal, i.e. best, component allocation one needs to compare all the feasible solutions $p_i$ from $\mathcal{P}$ and select the one with the smallest $w > 0$. Having many components ($n$) and computing units ($m$), the search space $|\mathcal{P}|=m^n$ tends to be a large number often hard (infeasible) to find, so SCALL incorporates a framework to reduce the search space and find sub--optimal good enough solution. 

\section{SCALL: Software component \\*allocator tool}

To implement SCALL, Eclipse was used, along with Eclipse Modeling Framework (EMF) and Graphical Modeling Project (GMF). This tool chain delivers a good platform for developing customized tools; EMF provides a modeling framework and code generation facilities, while GMF provides a set of generative components which create an infrastructure for development of graphical editors (Figure \ref{fig:screenshot}). 

\begin{figure}[ht]
    \centering
		\includegraphics[trim = 60mm 140mm 15mm 30mm, clip, width=0.95\columnwidth]{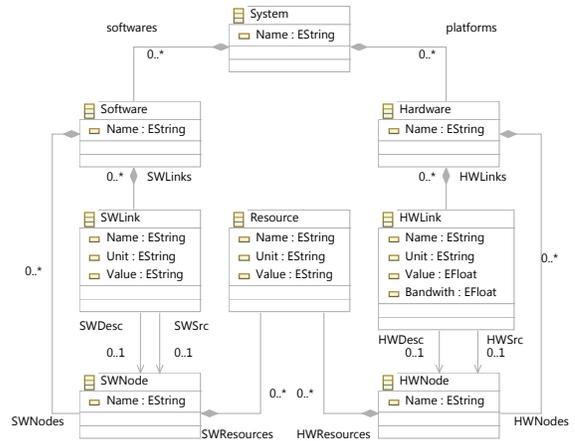}
    \caption{SCALL metamodel in Ecore notation}
    \label{fig:metamodel}
\end{figure}

SCALL consists of two main parts; a) \textsl{Eclipse based model} editor and b) \textsl{PyAllocator}.

 a) \textit{Eclipse based model editor} -- every model represented in SCALL editor corresponds to the metamodel depicted in Figure~\ref{fig:metamodel}. The metamodel uses a standard EMF Ecore notation. Unlike UML or its profiles (e.g. Marte) which could have been also used for SCALL, our metamodel uses a small number of straightforward concepts. Having a custom model, we can focus on the recognition of important concepts without a burden of large industry standard models and reason about them more effectively. In future, if they prove to be valuable, these concepts can be added to standard models. 
\\*
Eclipse based model editor is used to design the system. Figure~\ref{fig:metamodel} shows that the \texttt{System} consists of two main compartments: \texttt{Software} and \texttt{Hardware}. The \texttt{Software} is used to host a component model. It can consist of three main concepts; \texttt{SWNodes}, \texttt{SWLinks} and \texttt{Resources}. 
\texttt{SWNodes} represent software components which communicate via \texttt{SWLinks} and require some \texttt{Resources} to meet their execution specification. A \texttt{SWNode} can be associated with any number of links and resources, however all the \texttt{SWNode}s need to define values for all resources used in the model. \\*
Similarly, the computing platform is represented by \texttt{Hardware} which can host \texttt{HWNodes}, \texttt{HWLinks} and \texttt{Resources}. \texttt{HWNodes} represent (heterogeneous) computing units which communicate via \texttt{HWLinks} and provide a set of \texttt{Resources}. \texttt{HWLinks} represent physical communication media for which the current metamodel recognizes two attributes: bandwidth and communication cost (matrices $\mathcal{B}$ and $\mathcal{C}$). \texttt{Software} compartment hosts \texttt{SWNodes} with corresponding resource demand. For \texttt{SWLinks} through which components communicate, allocation model also uses communication intensity (currently a placeholder for future reference, e.g. number of function calls).

b) \textit{{PyAllocator}} -- a Python script which employs our multi--objective heuristic allocation method. Python was chosen for its simplicity in handling complex algorithms with rich and fast libraries (e.g. NumPy, SciPy) . From a model created in Eclipse, SCALL extracts previously described matrices which are necessary for making an allocation decision. Once all the matrices are collected, they are converted to JSON format and sent to the PyAllocator which performs the allocation. Using AHP, PyAllocator calculates weights for each resource (trade--off vector $\mathcal{F}$) and verifies the consistency of the users input. Finally, the solution is obtained by a genetic algorithm and sent to Eclipse based model editor to be displayed for the user. Detailed description of the decision model can be found in~\cite{Svogor2013}.

\subsection{Features}
 
To provide an insight about the software allocation on a heterogeneous platform and also to give a performance estimation, SCALL currently has the following features:

a) \textit{Model creation} --- the user can simultaneously model software and hardware architecture of the system. For software components, the user can add resources and model their communication channels. The same can be done for computing units, within the same view. \\*
b) \textit{Model visualization} --- enables easier model creation and better overview of the system. Instead of a classical approach of adding elements to a tree-like structure this approach enables drag--n--drop of model elements from a palette and a side--by--side view of both software and hardware architecture. \\*
c) \textit{Pairwise resource comparison} --- Having a complete model ready for analysis, the user needs to perform a manual pairwise resource comparison using AHP notation~\cite{saaty1994fundamentals} (i.e. determine which resources are more important). \\*
d) \textit{Software component allocation} --- SCALL uses the described process to perform a multi--objective allocation of software components on a heterogeneous platform. The procedure results in a (sub--)optimal solution. At any time, the user can re--run the allocation process to get alternative allocations. 
\\
We generated architectures with different number of components and computing units with solution space varying between $[5^3, 15^{10}]$. SCALL performed very well, i.e. the average time necessary for the  search spaces was (respectively) between $[0.59, 1.07]$ seconds while the exact solution found by full space search was up to $193$ seconds\footnote{2$\times$ Intel\textregistered Xeon\textregistered CPU E7-4830, 8GB Ram, Linux}. Given a genetic algorithm in the background, the exact solution was found for all tests and maximal error was $13$\% while, in average, SCALL was off by 3\%. 

\subsection{Tool usage scenario}

The following steps describe a typical usage scenario: a) determine software components, b) create a software architecture layout, c) determine computing units, d) create a hardware architecture layout, e) determine the necessary resources for all software components, f) determine the available resources on each computing unit, g) perform a pairwise comparison of resources and h) find the satisfactory allocation. 

\begin{figure}[ht]
    \centering
    \includegraphics[width=.99\linewidth]{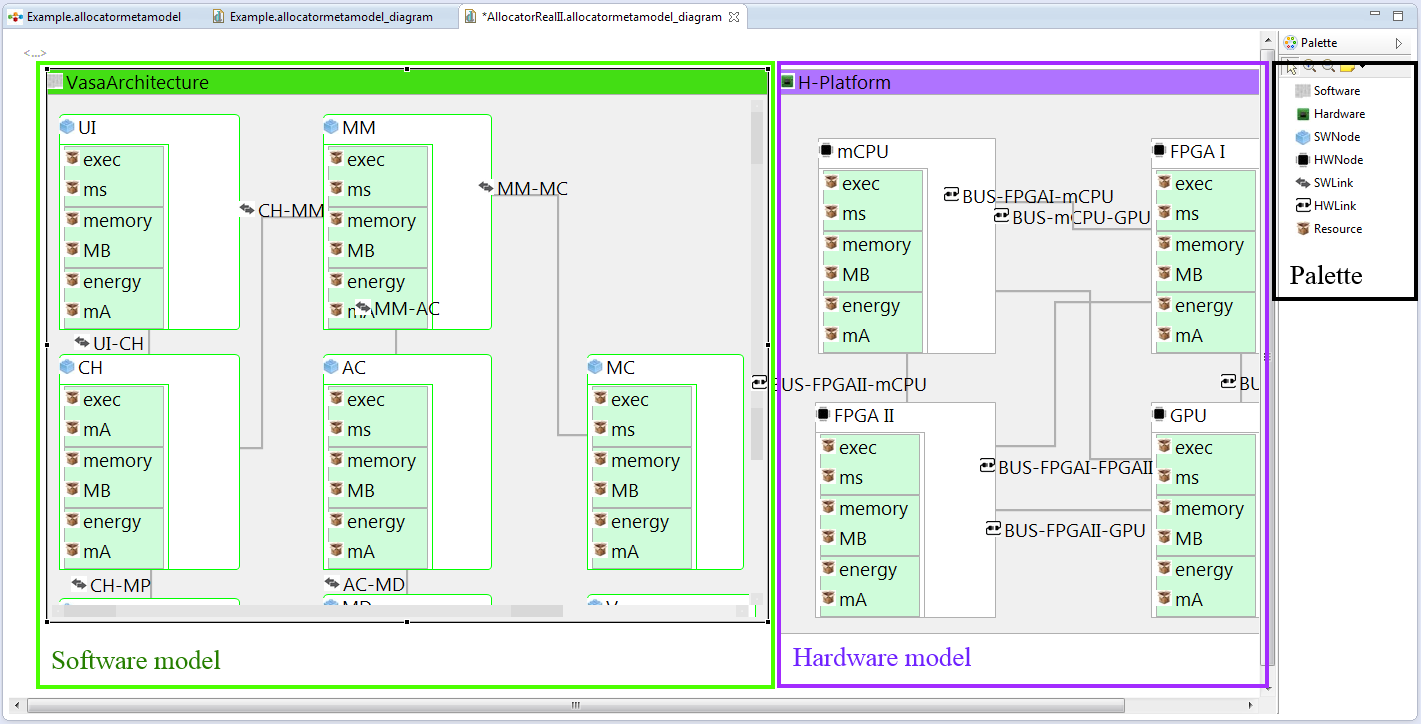}
    \caption{SCALL screen shot, showing software (left) and hardware (right) architecture side--by--side }
    \label{fig:screenshot}
\end{figure} 

In steps \textit{a)} and \textit{c)} the system architect needs to carefully consider functional and extra--functional properties to define components, interface contracts, computing units and communication channels. Links between components and computing units should be determined in steps \textit{b)} and \textit{d)} with consideration to bandwidth limitation. Steps \textit{e)} and  \textit{f)} ensure the quantification of resources in a determined measurement unit. All the same resources need to be quantified in same measurement unit (e.g. sample model in Figure \ref{fig:screenshot} considers memory in $\mathit{MB}$, computing time in $\mathit{\mu s}$  and energy in $\mathit{W}$). For step \textit{g)} the user needs to evaluate the significance of resources by a pairwise comparison. Higher mark for a resource gives more significance in allocation decision. In the final step, \textit{h)} the user initiates the allocation process which provides a user with the results.

\subsection{Deployment and installation}

All the described features of SCALL along with the software dependencies are deployed as an Eclipse plugin. The installation can be found at the web page: \url{http://thor.foi.hr/scall}. Currently, SCALL is available only for Windows.

\section{Related work}

Due to high dimensionality of modern enterprise systems and adaptive software systems which emerge from ubiquitous computing, some authors~\cite{da2011enterprise} argue that there is a lack of formal methods to cope with the complexity which result in deployment alternatives. This is also very important in early design phase because in that phase mistakes are the cheapest and the notion of performance can guide a system design. Authors mostly focus on two parameters which influence the system performance; communication delay~\cite{ccelik2013s} and computational delay~\cite{santinelli2010component,wang2004early}.
\\*
A tool by Malek et al. addresses architecture optimization problem by assessing and improving the quality of deployment in a particular scenario or class of scenarios. Having scenario based decision they can multiple possibly conflicting QoS dimensions~\cite{Malek2012}. Their tradeoff model utilizes users' preferences for the desired levels of QoS, find the most suitable deployment architecture.
\\*
In~\cite{ccelik2013s} authors present an allocation method to improve resource utilization and scheduling while keeping the decisions independent of platform implementation. The method determines the allocation using a genetic algorithm with an optimization function (heuristic algorithms are often used since allocation is a NP hard problem~\cite{ssaed2012metaheuristic}).
\\*
A recent systematic review by Aleti et al.~\cite{Aleti2013} shows that for optimizing software architecture researchers prefer to use custom modeling languages with allocation as a primary degree of freedom, followed by hardware replication and scheduling. However, current work largely focuses on allocating components and optimizing performance disregarding the emerging heterogeneous platforms. In this paper we are targeting high performance heterogeneous systems with various processing elements, e.g. CPU, GPU, FPGA.
 

\section{Potential Impact and \\*Future Work}
Allocating software components on computing units is an important challenge for the design decision makers in complex systems like automotive, avionics, manufacturing, etc. 
Our tool has a potential impact on: a) run--time and dynamic architectures b) (self--)adaptive systems and c) HSCD decision making. For a)--c) SCALL contributes to solving the challenge of design space exploration by supplying (sub--)~optimal design alternatives which system architects need for better decision making focusing on early design phase performance evaluation. In an early design phase the notion of system performance can be used to guide the system design towards an efficient utilization of resources. SCALL is currently an early prototype and will be extended in future. There are several pending tasks to implement in the near future: a) enable generating the application skeleton from the model, with performance parameters included as variables which will be measured on real systems, b) improve on theory to remove more constraints (e.g. cross-- computing unit components), c) refine the background algorithms through simulating performance evaluation with measured performance in existing systems.

\section*{Acknowledgment}

SCALL was partially supported by Swedish Foundation for Strategic Research grant (SSF) via project \textit{RALF3}. 
\vfill

%
\bibliographystyle{abbrv}
\bibliography{bibliography}  
%
%

\clearpage
\onecolumn

\appendix
\hfill
\\*
\textbf{Demonstration plan:}

\begin{itemize}
	\item \textit{Introduction and motivation }--- explain the issues related to software architecture planning and early design phases
	
	\item \textit{Background} --- briefly present the problem background and explain why component allocation on heterogeneous systems is an important problem.
	
	\item \textit{Live tool demo} --- we will apply the tool on a concrete problem related to allocation of software components on a heterogeneous system. In particular, it's an underwater autonomous vehicle (AUV) with 3 main computing units, \textsl{multicore CPU}, \textsl{GPU} and \textsl{FPGA} with 11 software components. First few components will be added manually, and the final example will be loaded from file.
	
	\item \textit{Conclusion and future work} --- conclude the presentation with current tool drawbacks, future implementation and potential impact.
	\item \textit{Q\&A}
\end{itemize}

\textbf{Tool availability and maturity}

SCALL is currently a prototype with some pending updates to be implemented in the near future. These include code a) skeleton generation, b) model update with additional optimization algorithms (simulated annealing), c) an alternative to AHP and d) re-validation of the underlying mathematical model in a real system (a robot with a heterogeneous platform, with different software allocations competing). A tool manual will also be available in the future, as suggested by users. 
\\*
Unfortunately, the tool is currently available only for Windows:

URL: \url{http://thor.foi.hr/scall} \\

CATEGORY: research \\

VIDEO: \url{https://youtu.be/dbsQZ36susw}\\

\end{document}